\documentclass[aps,prd,nofootinbib]{revtex4-2}

\usepackage{hyperref}
\usepackage{units}
\usepackage{epsfig}
\usepackage{pifont}
\usepackage{amsmath,amsfonts,amssymb,amsthm,mathrsfs}
\usepackage{graphicx,chngcntr,slashed}
\usepackage{cancel}
\usepackage{pstricks}
\usepackage{afterpage}
\usepackage{braket}
\usepackage{caption}
\usepackage{subcaption}
\usepackage{xcolor}
\usepackage{makecell}
\usepackage{physics}
\usepackage{multirow}
\usepackage{fancyhdr}
\usepackage{dsfont}

\newcommand{\dslash}[1]{\slash\!\!\! #1}
\newcommand{\kp}{k^{\prime}}
\newcommand{\pp}{p^{\prime}}
\newcommand{\be}{\begin{equation}}
\newcommand{\ee}{\end{equation}}
\newcommand{\bea}{\begin{eqnarray}}
\newcommand{\eea}{\end{eqnarray}}
\def\OMIT#1{}

\makeatletter
\def\section{\@startsection {section}{1}{\z@}{+3.0ex plus +1ex minus
  +.2ex}{2.3ex plus .2ex}{\large\bf\boldmath}}
\def\subsection{\@startsection{subsection}{2}{\z@}{+2.5ex plus +1ex
minus +.2ex}{1.5ex plus .2ex}{\normalsize\bf\boldmath}}
\def\subsubsection{\@startsection{subsubsection}{3}{\z@}{+3.25ex plus
 +1ex minus +.2ex}{1.5ex plus .2ex}{\normalsize\it}}

\begin{document}
\thispagestyle{empty}

\def\thefootnote{\fnsymbol{footnote}}

\title{Transverse spin asymmetries at the EIC as a probe of anomalous electric and magnetic dipole moments}

\author{Radja~Boughezal}
\affiliation{HEP Division, Argonne National Laboratory, Argonne, Illinois 60439, USA}
\author{Daniel de Florian}
\affiliation{International Center for Advanced Studies (ICAS), ICIFI and ECyT-UNSAM, 25 de Mayo y Francia, (1650) Buenos Aires, Argentina}
\author{Frank~Petriello}
\affiliation{HEP Division, Argonne National Laboratory, Argonne, Illinois 60439, USA}
\affiliation{Department of Physics \& Astronomy, Northwestern University, Evanston, Illinois 60208, USA}
\author{Werner Vogelsang}
\affiliation{Institute for Theoretical Physics, T\"ubingen University, Auf der Morgenstelle 14, 72076 T\"ubingen, Germany}

\begin{abstract}
  
  We show that inclusive single-spin asymmetries (SSAs) with transversely polarized protons or electrons at a future electron ion collider (EIC) are sensitive to new physics contributions to electroweak dipole operators of electrons and quarks. We use the Standard Model Effective Field Theory (SMEFT) to parameterize possible heavy new physics contributions to these couplings. We show that new physics scales at or beyond the TeV-scale can be probed assuming realistic EIC run parameters, and that the transverse spin asymmetries are sensitive to different combinations of the dipole couplings than other measurements such as anomalous magnetic or electric dipole moments. We also study the physics potential of SSAs at a possible future upgrade of the EIC to collide muons and protons. Measurements at such an upgrade could probe the same SMEFT parameters that explain the current anomaly in the muon anomalous magnetic moment, and could also improve current bounds on the muon electric dipole moment.
        
\end{abstract}

\def\thefootnote{\arabic{footnote}}
\setcounter{page}{0}
\setcounter{footnote}{0}

\maketitle


\section{Introduction}

The coming decade will see the construction of the Electron Ion Collider (EIC) at Brookhaven National Laboratory. The EIC will collide electrons with protons and nuclei at energies spanning the range between fixed-target scattering facilities and high energy colliders. It will provide orders of magnitude higher luminosity than HERA, the only electron-proton collider operated to date, and will also be the first lepton-ion collider with the ability to polarize both electron and light ion beams. The EIC was designed primarily to explore unresolved issues in QCD such as the composition of the proton spin in terms of its constituent quarks and gluons~\cite{Accardi:2012qut}. The EIC also has the potential to explore possible extensions of the Standard Model (SM). The possibility of polarizing both beams provides novel opportunities for probes of new physics complementary to those possible at the LHC. For example, measurements of parity-violating longitudinal spin asymmetries can constrain combinations of four-fermion operators orthogonal to the combinations probed at the LHC~\cite{Boughezal:2020uwq,Boughezal:2022pmb}. 

Choosing transverse polarization for the electron or light ion beams at the EIC will enable measurements of beam and target transverse single-spin asymmetries (SSAs).  The most basic transverse SSAs are obtained for inclusive deep-inelastic scattering (DIS) and have been 
studied previously within the SM in Refs.~\cite{Afanasev:2007ii,Metz:2006pe,Schlegel:2009pw,Metz:2012ui,Schlegel:2012ve}. As was shown already in~\cite{Christ:1966zz}, these asymmetries vanish for purely electromagnetic scattering in the one-photon 
exchange approximation. Beyond that, the spin-dependent numerator of the SSA is suppressed by both a power of the fine structure
constant $\alpha$ and a factor of $m/Q$, where $m$ is the mass of the polarized particle, and $Q$ is the deep-inelastic scattering (DIS) momentum transfer. 
Although we will identify in this paper a new tree-level source of transverse SSAs in the SM not previously discussed in the literature, 
the upshot is that the SM predicts that the inclusive transverse SSAs are strongly suppressed, 
with target asymmetries that are numerically of the order $10^{-4}$ and beam asymmetries of the order $10^{-7}$. These extremely small SM values, combined with the expected excellent experimental precision of the EIC, make these asymmetries a potentially powerful probe of new physics that does not contain the suppression factors present in the SM.


In this manuscript we study the sensitivity of transverse SSAs to heavy new physics. We use the SM Effective Field Theory (SMEFT) to parameterize physics beyond the SM~\cite{Buchmuller:1985jz,Arzt:1994gp,Grzadkowski:2010es}. The SMEFT is formed by adding higher-dimensional operators to the SM Lagrangian that are consistent with the SM gauge symmetries and formed only from SM fields. The SMEFT encapsulates a broad swath of new physics models, making it easier to simultaneously study numerous theories without focusing on details of the their ultraviolet completions. We show that measurements of the SSAs at the EIC are sensitive probes of fermion dipole couplings to photons and $Z$-bosons. In particular, transverse beam SSAs are sensitive to dipole couplings of electrons, while target SSAs are sensitive to quark dipole couplings. We find that within the SMEFT both real and imaginary parts of the dipole couplings can contribute to the transverse SSAs. Their effects can be disentangled through their angular dependence. Other experimental probes of these couplings include anomalous magnetic moments, electric dipole moments, and Drell-Yan measurements at the LHC. Transverse SSAs probe different parameter combinations than these other searches and are therefore complementary to these other measurements. We show that new physics at the TeV scale could be studied at the EIC. In addition to our SMEFT analysis we identify a new source of transverse SSAs in the SM that will provide the dominant contribution at EIC energies. One possible upgrade discussed for the EIC is the replacement of the electron beam with a high energy muon beam. This could serve as a first step toward a high energy muon-muon collider. We show that measurements of SSAs at a muon-ion collider could probe parameter space relevant for the muon $g-2$ anomaly, and could also improve upon current bounds on the muon electric dipole moment.

Our manuscript is organized as follows. We review the calculation of transverse SSAs in the SM in Section~\ref{sec:SMreview}. In this section we point out a new mechanism for generating these asymmetries in the SM that has not been discussed previously, and that will be the dominant mechanism at the EIC. In Section~\ref{sec:SMEFT} we discuss aspects of the SMEFT relevant for our calculation, and discuss the calculation of transverse SSAs within the SMEFT. We present numerical results for the transverse SSAs in the SMEFT in Section~\ref{sec:numerics}. We present simple estimates of the anticipated experimental error at the EIC that indicate that TeV new physics scales should be accessible with transverse SSA measurements. In Section~\ref{sec:other} we briefly discuss other experimental probes of the parameter space and demonstrate that EIC measurements will be complementary to them. We discuss transverse SSAs at a muon-ion collider in Section~\ref{sec:muon}. We show that such measurements could probe parameter space relevant for the current discrepancy between theory and experiment in the muon anomalous magnetic moment, and could improve current bounds on the muon electric dipole moment. Finally, we conclude in Section~\ref{sec:conc}.

\section{Transverse SSA in the SM}
\label{sec:SMreview}

We revisit here the SM calculation of the transverse SSA in the inclusive DIS process $e(k)+p(P) \to e(\kp)+X$. Assuming that both initial beams are along the $\hat{z}$-axis, we can write the transverse spin vector of either the electron or the proton as
\be
    S_T^{\mu} = \left(0,\text{cos}(\phi),\text{sin}(\phi),0 \right)
\ee
where $\phi$ denotes the angle between the transverse spin and the direction of the outgoing lepton in the transverse plane. The asymmetry is then defined as the difference of the cross sections for positive and negative $S_T$ divided by their sum. If the initial electron is polarized it is called a beam SSA, while if the initial proton is polarized it is referred to as a target SSA. For instance, in the case of the beam asymmetry the expression takes the form
\be
A_{TU} = \frac{\sigma(e^\uparrow)-\sigma(e^\downarrow)}{\sigma(e^\uparrow)+\sigma(e^\downarrow)},
\ee
where we have used up and down arrow superscripts to denote positive and negative $S_T$. A similar expression holds for the target asymmetry with the replacement of polarized electrons with polarized protons.

In the SM neither SSA is generated by QED at tree level~\cite{Christ:1966zz}. The leading QED contribution comes from two-photon exchange and is therefore suppressed by a power of $\alpha$, the fine structure constant. Furthermore, the calculation of the two-photon exchange contribution requires a mass insertion along either the electron line (for the beam SSA) or the quark line (for the target SSA computed in the parton model). The simplest way to see this is to note that the spin projector for a massive fermion with transverse spin $S_T$ can be written as
\be
u(p)\bar{u}(p) = \frac{1}{2}(\dslash{p}+m)(1+\gamma_5 \dslash{S_T}).
\label{eq:transverseproj}
\ee
The terms dependent on $S_T$ change the numbers of gamma matrices appearing from even to odd or vice versa, therefore changing the number of mass insertions required to have a non-zero trace when computing a squared amplitude. The two-photon exchange contribution to the SSA can be shown to depend only on the structure 
\be
\epsilon_{\mu\nu\rho\sigma}k^{\mu}k^{\prime\nu}P^{\rho}S_T^{\sigma}.
\ee
This structure is naively time-reversal odd~\cite{Hagiwara:1982cq}, and requires a complex phase in order to contribute to an observable. Combining these two effects leads to an $\alpha \times m/Q$ suppression.

The calculation of both beam and target asymmetries in QED has been considered previously~\cite{Afanasev:2007ii,Metz:2006pe,Schlegel:2009pw,Metz:2012ui,Schlegel:2012ve}. The result for the beam asymmetry can be written as~\cite{Schlegel:2009pw} 
\be
A_{TU}^{\gamma\gamma}(\phi) = \alpha \frac{m_l}{2Q} \text{sin}(\phi) \frac{y^2\sqrt{1-y}}{1-y+y^2/2}\frac{\sum_q Q_q^3 f_q(x)}{\sum_q Q_q^2 f_q(x)}.
\ee
Here, $f_q$ denotes the parton distribution function (PDF) of quark $q$, $Q_q$ denotes its electric charge, $x$ denotes Bjorken-$x$, $Q$ is the usual DIS momentum transfer and $y$ is the  DIS inelasticity parameter. Since the EIC will operate at relatively high momentum transfers, the leading-twist approximate is the appropriate language here. The calculation of the target SSA is more intricate. The same two-photon exchange contribution gives~\cite{Metz:2006pe}
\be
A_{UT}^{\gamma\gamma}(\phi) = \alpha \frac{M}{2Q} \text{sin}(\phi) \frac{y\sqrt{1-y}}{1-y+y^2/2}\left( \text{ln}\left(\frac{Q^2}{\lambda^2}\right)+\text{finite}\right)\frac{\sum_q Q_q^3 g^T_q(x)}{\sum_q Q_q^2 f_q(x)}
\ee
where $g^T_q$ denotes a higher-twist PDF, and $M$ is the target nucleon mass. $\lambda$ denotes a small photon mass that regulates an infrared divergence appearing in the calculation, whose presence clearly indicates the inadequacy of the parton model in describing this result. 
As was later shown~\cite{Schlegel:2012ve}, the dependence on $\lambda$ cancels once one takes into account 
quark transverse motion and mass effects, as well as contributions from $qgq$ correlations in the nucleon. In this way, a well-defined
finite answer for $A_{UT}^{\gamma\gamma}$ is obtained. In addition, there could also be two-photon exchange contributions
for which the photons couple to two different quark lines, turning out to be sensitive to $q\gamma q$ correlation functions~\cite{Metz:2012ui}.
In any case, simple model calculations give an asymmetry in the range $A_{UT}^{SM} \sim 10^{-4} - 10^{-3}$~\cite{Afanasev:2007ii,Metz:2012ui}.

At the higher momentum transfers relevant for the EIC we must also include SM contributions mediated by the $Z$-boson, which have not been previously considered. The $Z$-boson contribution grows as $Q^2/M_Z^2$ for moderate values of $Q^2$. The leading contribution comes from interference between photon and $Z$
exchange, and for the beam asymmetry can be written as
\be
A_{TU}^Z(\phi) = \frac{2}{s_W^2 c_W^2} \frac{m_l Q}{M_Z^2} \frac{y\sqrt{1-y}}{1-y+y^2/2} \text{cos}(\phi) \frac{\sum_q Q_q f_q(x)  \left[ g_{al} g_{vq} (1-y) + g_{vl} g_{aq} y\right] }{\sum_q Q_q^2 f_q(x)}.
\ee
Here, $s_W$ and $c_W$ respectively denote the sine and cosine of the weak mixing angle, while the vector and axial couplings of the fermions are
\be
g_{vf} = \frac{T_3^f}{2} - Q_f s_W^2,\;\;\; g_a = -\frac{T_3^f}{2}.
\ee
For simplicity of presentation we have expanded this result to leading order in the ratio $Q^2/M_Z^2$. The expressions for the anti-quark channels can be obtained by taking $g_{aq} \to -g_{aq}$. Our numerical results include all partonic channels as well as the full $Q^2$ dependence and the self-interference of the $Z$-exchange diagram, both of which are numerically sub-dominant. We note that the $Z$-boson exchange depends on the dot product $k^{\prime} \cdot S_T$, and therefore has a different dependence on the angle $\phi$. We also note that each term in this expression depends linearly on an axial coupling of the $Z$-boson to fermions, indicating that this is a parity-violating effect. The full asymmetry in the SM is the sum of the two-photon contribution and the one involving the $Z$ boson.

\begin{figure}[h!]
\centering
\includegraphics[width=0.7\textwidth]{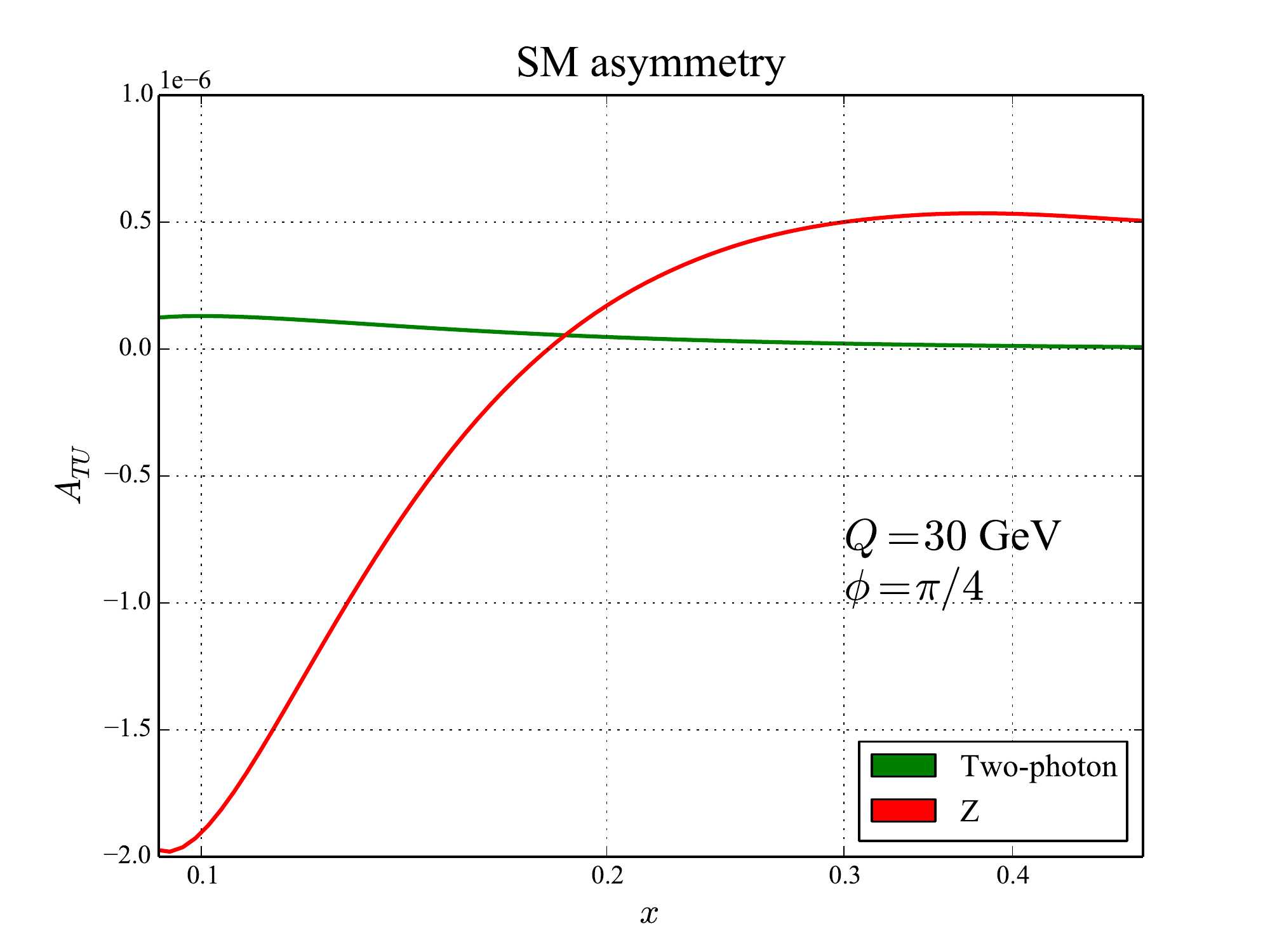}
\caption{The magnitudes of the two-photon and $Z$-exchange contributions to the SM asymmetry for $y=1/2$ as a function of momentum transfer. The $y$-axis is in units of $10^{-6}$.
\label{fig:ASM}}
\end{figure}

To show the relative size of these two contributions we plot them assuming $\phi=\pi/4$ in Fig.~\ref{fig:ASM} as a function of $x$ assuming the representative momentum transfer $Q=30$ GeV. We note that for this choice of angle both mechanisms contribute. For most values of $Q$ relevant for a higher-energy BSM analysis the $Z$-boson exchange dominates. Thanks to their different dependence on $\phi$ one may in principle disentangle the two contributions by taking moments of the asymmetry weighted with $\sin(\phi)$ or $\cos(\phi)$, respectively. 

A similar contribution from $Z$-boson exchange occurs for the target asymmetry. We can calculate it to be\footnote{
We note that in this expression we only keep the contributions by the leading-twist transversity PDF. As is evident from the
explicit proportionality to the quark mass $m_q$, the asymmetry is power-suppressed. As a result, there will be additional contributions 
associated with higher-twist PDFs. Using the techniques presented in~\cite{Kanazawa:2015ajw} we find the replacements
$(1-y)m_qh_q\to (1-y)\big(m_qh_q+M xg_q^T-Mg_{1T,q}^{(1)}\big)-Mg_{1T,q}^{(1)}$ in the $g_{aq} g_{vl}$ part of 
the asymmetry, and $m_qh_q y\to y\big(m_qh_q+M xg_q^T-Mg_{1T,q}^{(1)}\big)-Mxg_q^T$ in the $g_{vq} g_{al}$ part.
Here, as before, $g_q^T$ denotes a higher-twist PDF and $g_{1T,q}^{(1)}$ is the second moment of a transverse-momentum
dependent PDF. For our present analysis that aims at an order-of-magnitude estimate of the asymmetry, we ignore these additional contributions.
Given that even less is known about the $g_q^T$ and $g_{1T,q}^{(1)}$ distributions than about transversity, this appears
justified.}
\be
A_{UT}^Z(\phi) = -\frac{2}{s_W^2 c_W^2} \frac{m_q Q}{M_Z^2} \frac{y\sqrt{1-y}}{1-y+y^2/2} \text{cos}(\phi) \frac{\sum_q Q_q h_q(x)  \left[ g_{aq} g_{vl} (1-y) + g_{vq} g_{al} y\right] }{\sum_q Q_q^2 f_q(x)}.
\ee
The function $h_q$ denotes the twist-2 quark transversity distribution~\cite{Barone:2001sp}. These functions are currently still rather poorly known, although some extractions from data have been presented~\cite{Radici:2015mwa,Kang:2015msa,Lin:2017stx,Radici:2018iag}. Transversity distributions satisfy the Soffer inequality~\cite{Soffer:1994ww}
\be
2| h(x,\mu)| \leq f(x,\mu)+\Delta f(x,\mu)
\label{eq:soffer}
\ee
where $\Delta f$ is the helicity-dependent PDF. We will discuss later various model estimates for the transversity distributions. For the lighter quarks, it has been suggested that the quark mass appearing in this expression should be interpreted as a vacuum expectation value in the presence of non-perturbative vacuum fields leading to a constituent mass $m_q \sim M_{proton}/3$~\cite{Afanasev:2007ii}. We note that the integral of the transversity distribution is related to the tensor charge that appears when converting quark electric dipole moments (EDMs) to nucleon EDMs~\cite{Liu:2017olr}. In a later section we only consider the muon EDM that can be probed by measurements of the beam asymmetry, and therefore the tensor charge does not enter our analysis.

\section{Transverse SSA in the SMEFT}
\label{sec:SMEFT}

In this section we review aspects of the SMEFT needed in our study, and discuss the leading contributions to both beam and target SSAs. The SMEFT is an effective field theory extension of the SM that includes terms
suppressed by a high energy scale $\Lambda$. Above this scale the ultraviolet completion of the EFT 
becomes important, and new particles beyond the SM appear.  In our study we keep terms through dimension-6 in the $1/\Lambda$ expansion, and 
ignore operators of odd-dimension which violate lepton number. Our Lagrangian becomes~\cite{Buchmuller:1985jz,Arzt:1994gp,Grzadkowski:2010es}
\begin{equation}
{\cal L} = {\cal L}_{SM}+ \sum_i C^{(6)}_{i} {\cal
  O}^{(6)}_{i} + \ldots,
\end{equation}
where the ellipsis denotes operators of higher dimensions. The Wilson coefficients $C^{(6)}_{i}$ have dimensions of inverse energy squared. Cross sections computed through linear order in the Wilson coefficients will have interferences between dimension-6 operators and the SM.

We will look for contributions to the transverse SSAs in the SMEFT that are not suppressed like the SM terms. In order to get a contribution from a SMEFT operator not subject to the electron or quark mass suppression present in the SM, there must be a chirality violation coming from a new source within the SMEFT. Consideration of the possible operators at dimension-6 reveals the following categories that can potentially lead to such an effect: scalar or tensor four-fermion operators, new Higgs-boson interactions not proportional to fermion masses, and dipole operators of fermions. Only the third category contributes without an explicit mass suppression at the dimension-6 level. To illustrate this finding we will discuss the contribution of the scalar and tensor operators in detail. There are three such operators which we write below, suppressing generation indices but keeping SU(2) indices:
\bea
{\cal O}_{ledq} &=& (\bar{l}^j e)(\bar{d}q^j) , \nonumber \\
{\cal O}^{(1)}_{lequ} &=& (\bar{l}^j e)\epsilon_{jk}(\bar{q}^ku) , \nonumber \\
{\cal O}^{(3)}_{lequ} &=& (\bar{l}^j \sigma^{\mu\nu} e)\epsilon_{jk}(\bar{q}^k \sigma_{\mu\nu} u) .
\eea
$l$ denotes the left-handed SU(2) lepton doublet, $e$ denotes the right-handed SU(2) electron singlet, $q$ represents the left-handed SU(2) quark doublet, and $u$, $d$ denote the right-handed singlet quarks. We can illustrate the main points of the calculation using ${\cal O}_{ledq}$ as an example. All Feynman rules for these operators can be found in Ref.~\cite{Dedes:2017zog}. The contribution to the parton-level amplitude for the process $e(k)+q(p) \to e(\kp)+q(\pp)$ coming from ${\cal O}_{ledq}$ can be written as
\be
{\cal M} =  C_{ledq}^{*} [\bar{u}(\kp) P_L u(k)][ \bar{u}(\pp) P_R u(p)] + C_{ledq} [\bar{u}(\kp) P_R u(k)][ \bar{u}(\pp) P_L u(p)] 
\ee 
where $P_{L,R}=\frac{1}{2}(\mathds{1}\mp\gamma^5)$. When interfered with the SM tree-level amplitude and summed over spins assuming the transverse spin for the intitial electron shown in Eq.~(\ref{eq:transverseproj}), all terms contain the trace structure
\be
\text{Tr}\left[ (\dslash{\pp}+m_q) P_R (\dslash{p} +m_q) \gamma^{\mu} \right].
\ee
This has an odd number of $\gamma$ matrices and vanishes unless there is a mass insertion along the quark line. The same argument holds for the lepton line in the case of the target asymmetry. This is also mass-suppressed if we consider the dimension-6 squared contribution. In the massless limit for the beam asymmetry this contribution has the trace structure
\be
|C_{ledq}|^2 \text{Tr}\left[ \dslash{\kp}P_L\dslash{k}(1+\gamma_5 \dslash{S}_T) P_R \right] \text{Tr}\left[ \dslash{\pp} P_R \dslash{p} P_L \right].
\ee
All terms with the $S_T$ dependence have an odd number of $\gamma$ matrices in the trace. Helicity flips are needed on both the lepton and quark lines. Similar arguments hold for the following operators which mediate Higgs  ($\varphi$) exchange corrections:  
\bea
{\cal O}_{e\varphi} &=& (\varphi^{\dagger} \varphi) (\bar{l}e \varphi), \nonumber \\
{\cal O}_{u\varphi} &=& (\varphi^{\dagger} \varphi) (\bar{q}u \tilde{\varphi}), \nonumber \\
{\cal O}_{d\varphi} &=& (\varphi^{\dagger} \varphi) (\bar{q}d \varphi).
\eea

These arguments leave the following dipole operators as potentially enhanced contributions to the transverse SSAs:
\bea
{\cal O}_{eW} &=& (\bar{l}\sigma^{\mu\nu}e) \tau^I \varphi W^I_{\mu\nu},\nonumber \\
{\cal O}_{eB} &=& (\bar{l}\sigma^{\mu\nu}e) \varphi B_{\mu\nu},\nonumber \\
{\cal O}_{uW} &=& (\bar{q}\sigma^{\mu\nu}u) \tau^I \varphi W^I_{\mu\nu},\nonumber \\
{\cal O}_{uB} &=& (\bar{q}\sigma^{\mu\nu}u) \varphi B_{\mu\nu},\nonumber \\
{\cal O}_{dW} &=& (\bar{q}\sigma^{\mu\nu}d) \tau^I \varphi W^I_{\mu\nu},\nonumber \\
{\cal O}_{dB} &=& (\bar{q}\sigma^{\mu\nu}d) \varphi B_{\mu\nu}.
\eea
Here, $W^I$ and $B$ are the field strength tensors of the SM SU(2) and U(1) gauge groups, and the $\tau^I$ denote the Pauli matrices. We have written down these operators assuming first generation fermions. Identical operators with different Wilson coefficients can be written down for other fermion generations. The operators ${\cal O}_{eW}$ and ${\cal O}_{eB}$ provide the chirality flip needed for a non-vanishing beam SSA. The other operators lead to non-vanishing target SSAs. Whether the Wilson coefficients associated with these operators are proportional to the masses of the corresponding fermions depends on the details of the ultraviolet theory that lead to these operators. In the presence of new mass scales in the high-energy theory these parameters can be uncorrelated with the electron or quark masses. In this paper we make no assumptions about the underlying UV theory and treat the Wilson coefficients as free parameters. To leading order in the $Q^2/M_Z^2$ expansion we can write the SMEFT-induced correction to the beam asymmetry as
\be
\Delta A_{TU}(\phi) = \frac{g_Z}{2\pi\alpha} \frac{Q^3}{M_Z^2} \frac{y\sqrt{1-y}}{1-y+\frac{y^2}{2}} \frac{\sum_q Q_q f_q(x) \left\{ g_{aq} \text{Re}[C_{eZ} e^{-i\phi}]
	-\frac{\text{Re}[C_{e\gamma} e^{-i\phi}]}{s_W c_W} \left[ g_{vq} g_{al} (1-2/y)-g_{aq} g_{vl} \right]\right\} }{\sum_q Q_q^2 f_q(x)}
\label{ATUsmeft}
\ee
where $g_Z$ is related to the electric charge and weak mixing angle according to $g_Z=e/(s_Wc_W)$.
For simplicity of presentation, we have again shown the expression expanded to leading order in $Q^2/M_Z^2$; 
in our numerical results we include the full $Q^2$ dependence. The results for the anti-quark channels can be obtained from these results in the same way as for the SM anti-quark expressions. We have written the result in terms of the linear combination:
\bea
C_{e\gamma} &=& \frac{v}{\sqrt{2}} \left[ -s_W C_{eW} +c_W C_{eB} \right], \nonumber \\
C_{eZ} &=& \frac{v}{\sqrt{2}} \left[ -c_W C_{eW} -s_W C_{eB} \right]
\eea
which have dimensions of inverse energy. We note that the combination $C_{e\gamma}$ is the Wilson coefficient of the operator ${\cal O}_{e\gamma} = \bar{e}_L \sigma^{\mu\nu} e_R F_{\mu\nu}$ that describes the anomalous magnetic and electric dipole moments of the electron below the EW symmetry breaking scale. The particular combinations of Wilson coefficients and angle that appear in the result can be expressed in terms of the combinations $\text{Im}[C_{ei}] \text{sin}(\phi)$ and $\text{Re}[C_{ei}] \text{cos}(\phi)$. This dependence on the angle indicates that the real and imaginary parts of the Wilson coefficient can be separately determined via angular measurements. The asymmetry in Eq.~(\ref{ATUsmeft})
grows with momentum transfer, making the EIC an excellent facility to search for them. A similar expression holds for the target SSA. We show the expression below:
\be
\Delta A_{UT}(\phi) = \frac{g_Z}{2\pi\alpha} \frac{Q^3}{M_Z^2} \frac{y\sqrt{1-y}}{1-y+\frac{y^2}{2}} \frac{\sum_q Q_q h_q(x) \left\{ -g_{al} \text{Re}[C_{qZ} e^{i\phi}]
	-\frac{\text{Re}[C_{q\gamma} e^{i\phi}]}{s_W c_W} \left[ g_{vl} g_{aq} (1-2/y)-g_{al} g_{vq} \right]\right\} }{\sum_q Q_q^2 f_q(x)}.
\label{AUTsmeft}
\ee
We have again shown the expression expanded to leading order in $Q^2/M_Z^2$. 

\section{Numerical analysis of the SMEFT asymmetry}
\label{sec:numerics}

We study here numerical predictions for the SMEFT-induced corrections to the various asymmetries at a future EIC. To isolate either the real or imaginary parts of the Wilson coefficients experimentally we can weight events by the value of $\phi$ determined experimentally by forming an integrated asymmetry:
\be
A^w_{TU} = \int_{0}^{2\pi} d\phi \, w(\phi) \, A_{TU}(\phi).
\ee
For example, the weight function $w=\text{cos}(\phi)$ projects out the real part of the Wilson coefficient in 
$\Delta A_{TU}(\phi)$, while the $\text{sin}(\phi)$ proportional to the imaginary part integrates to zero. This angle is determined by the directions of the initial-state transverse spin and the final-state lepton direction. We expect that this angular quantity can be accurately measured at a future EIC.

\begin{figure}[h!]
\centering
\includegraphics[width=0.49\textwidth]{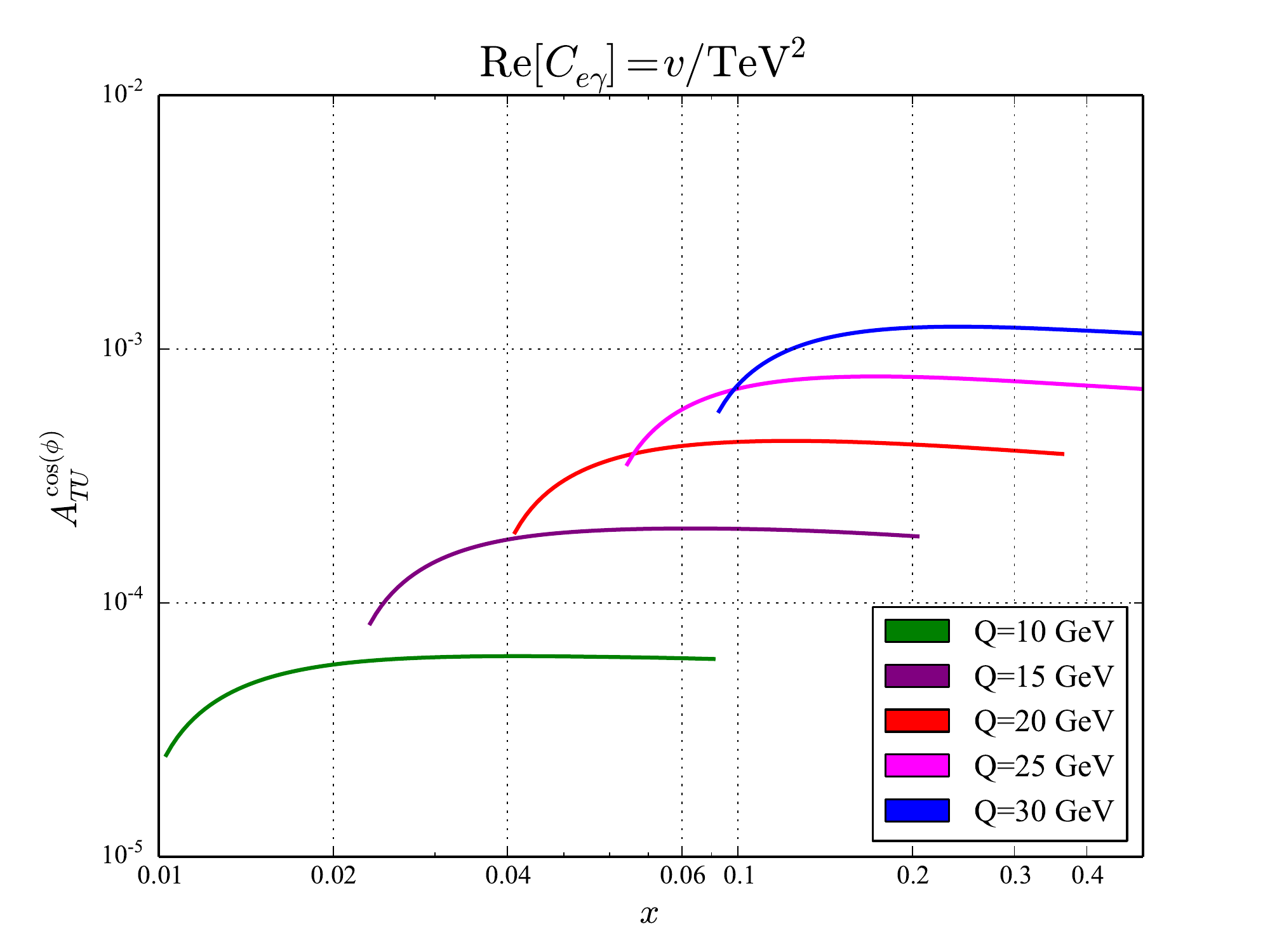}
\includegraphics[width=0.49\textwidth]{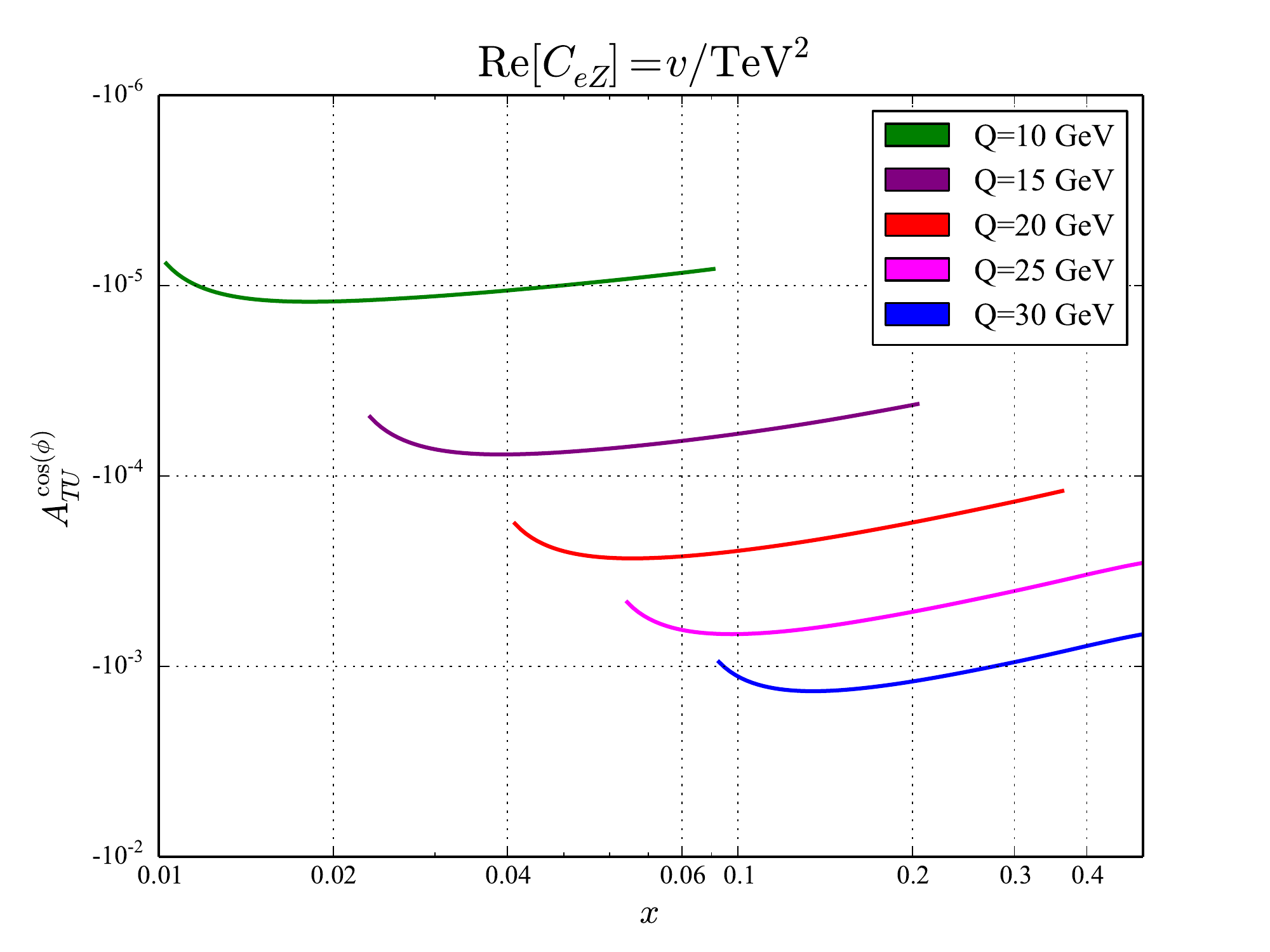}
\caption{The SMEFT contribution to the asymmetry for the real part of $C_{e\gamma}$ (left panel) and $C_{eZ}$ (right panel) for representative values of $Q$ as a function of Bjorken-$x$. 
\label{fig:ASMEFT}}
\end{figure}

We show in Fig.~\ref{fig:ASMEFT} the SMEFT contribution to the asymmetry for the real parts of $C_{e\gamma}$ and $C_{eZ}$ for representative values of $Q$ as a function of Bjorken-$x$. We have chosen $\text{Re}[C_{ei}]=v/\text{TeV}^2$ in both cases, where $v$ is the Higgs vacuum expectation value. This scaling is consistent with the expectation that these coefficients are generated by dimension-6 operators in the SMEFT at the TeV-scale. We assume $10 \, \text{GeV} \times 275 \, \text{GeV}$ collisions for a center of mass energy $\sqrt{s}=105$ GeV. Although this is not the highest possible energy at a future EIC, it is expected that this configuration will lead to the highest integrated luminosity. Previous studies have shown that maximizing the integrated luminosity leads to higher sensitivity to SMEFT parameters than a slight increase of energy~\cite{Boughezal:2022pmb}. We have imposed the inelasticity cuts $0.1<y<0.9$ in producing these results. As expected from their $Q^3$ functional dependence at intermediate momentum transfers the SMEFT asymmetries increase quickly with energy,  exceeding the $10^{-3}$ level for $Q>30$ GeV. The imaginary parts of the Wilson coefficients leads to identical integrated asymmetries after the appropriate change in the weight function. This is due to the structure of the asymmetry, which depends on the combination $\text{Re}[C_{ei} e^{-i\phi}]=\text{Re}[C_{ei}]\text{cos}(\phi)+\text{Im}[C_{ei}]\text{sin}(\phi)$ with $i=\gamma,Z$. We note that the photon and $Z$ dipole contributions come with opposite sign.

\begin{figure}[h!]
\centering
\includegraphics[width=0.7\textwidth]{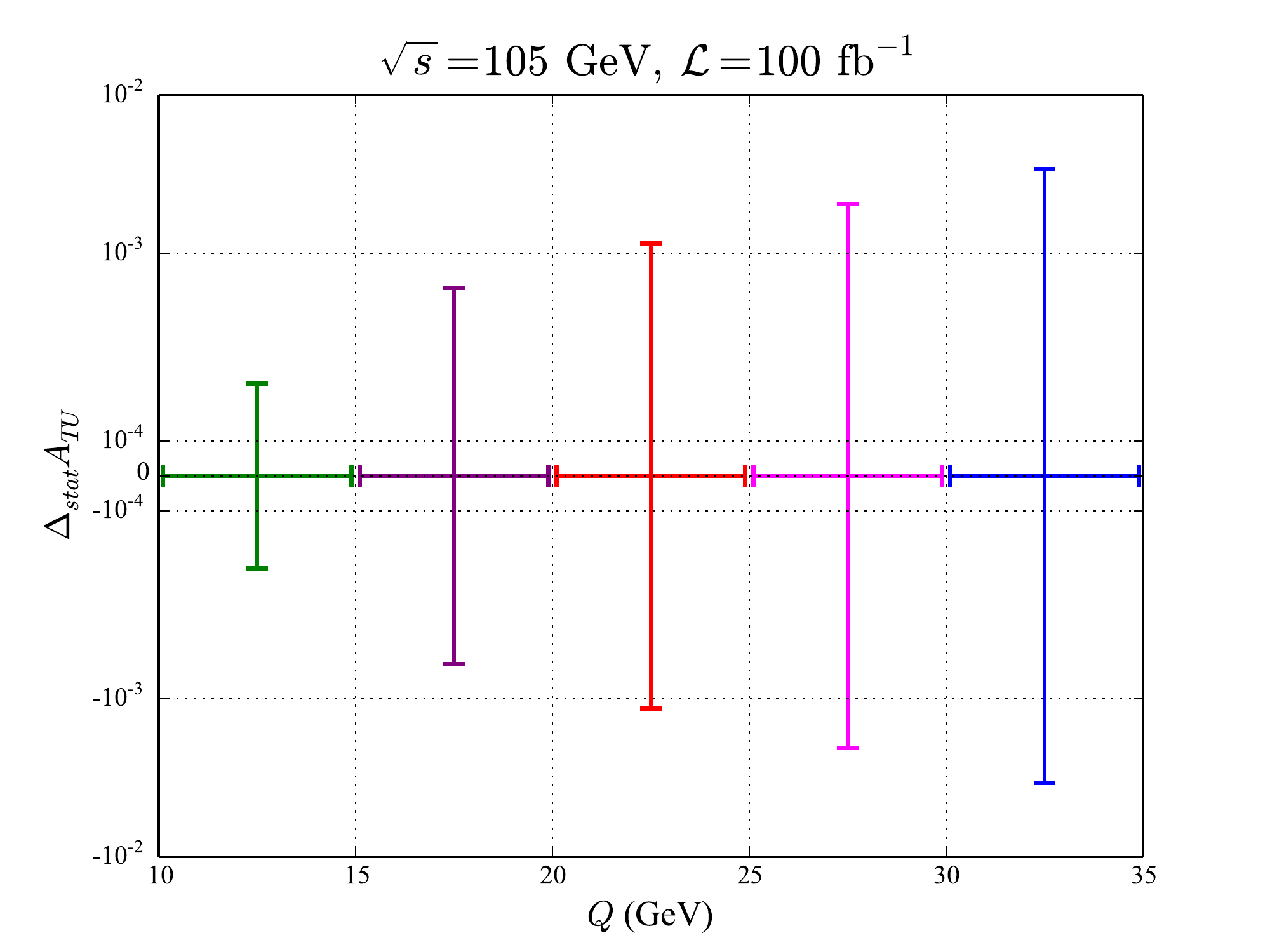}
\caption{The estimated statistical uncertainty on the asymmetry at a future EIC for $Q$ bins ranging from 10 to 35 GeV. Bins of width 5 GeV are assumed, and Bjorken-$x$ is integrated over subject to the constraint $x<0.5$. 
\label{fig:error}}
\end{figure}

Although an experimental simulation of this asymmetry at a future EIC is beyond the scope of this analysis, we briefly discuss the experimental reconstruction of this asymmetry and estimate the precision achievable at a future EIC. We denote the number of measured events with positive and negative transverse polarization as $N_{\uparrow\downarrow}$. Setting the achievable magnitude of transverse polarization as $|P_T|$, we can solve to find
\be
A_{TU} = \frac{1}{|P_T|} \frac{\int_{0}^{2\pi} d\phi \, \text{cos}(\phi) \left[ N_{\uparrow}(\phi)-N_{\downarrow}(\phi)\right]}{ N_{\uparrow}+ N_{\downarrow}}.
\ee
In the limit that the asymmetry is much less than one, and assuming that the only errors come from the polarization and the statistical error, we can write the uncertainty in the asymmetry as the sum in quadrature of two pieces:
\be
\delta A_{TU} = \frac{\delta P_T}{|P_T|} A_{TU} \oplus \frac{1}{|P_T|\sqrt{N_{\uparrow}+ N_{\downarrow}}}.
\ee
Since the error in the determination of the polarization fraction is expected to be at the percent level, the first term in this expression should lead to a small relative error on the asymmetry measurement. The potentially limiting uncertainty is the statistical uncertainty represented by the second term. We evaluate this by calculating the total number of events expected in the SM for various bins of momentum transfer, integrated over Bjorken-$x$ subject to the constraint $x<0.5$. We assume 100 fb$^{-1}$ of integrated luminosity at $\sqrt{s}=105$ GeV, a realistic operating point used in previous EIC studies~\cite{Boughezal:2022pmb}. The results are shown in Fig.~\ref{fig:error}. The statistical uncertainty is at or below the $10^{-3}$ level for $Q<25$ GeV, commensurate with the size of the beam SSA. We note that integrating the asymmetry over Bjorken-$x$ would increase the magnitude of the asymmetries presented in Fig.~\ref{fig:ASMEFT}. Although this simple estimate does not replace a realistic experimental analysis, it indicates that new physics scales at the TeV level can be probed at a future EIC.

\begin{figure}[h!]
\centering
\includegraphics[width=0.7\textwidth]{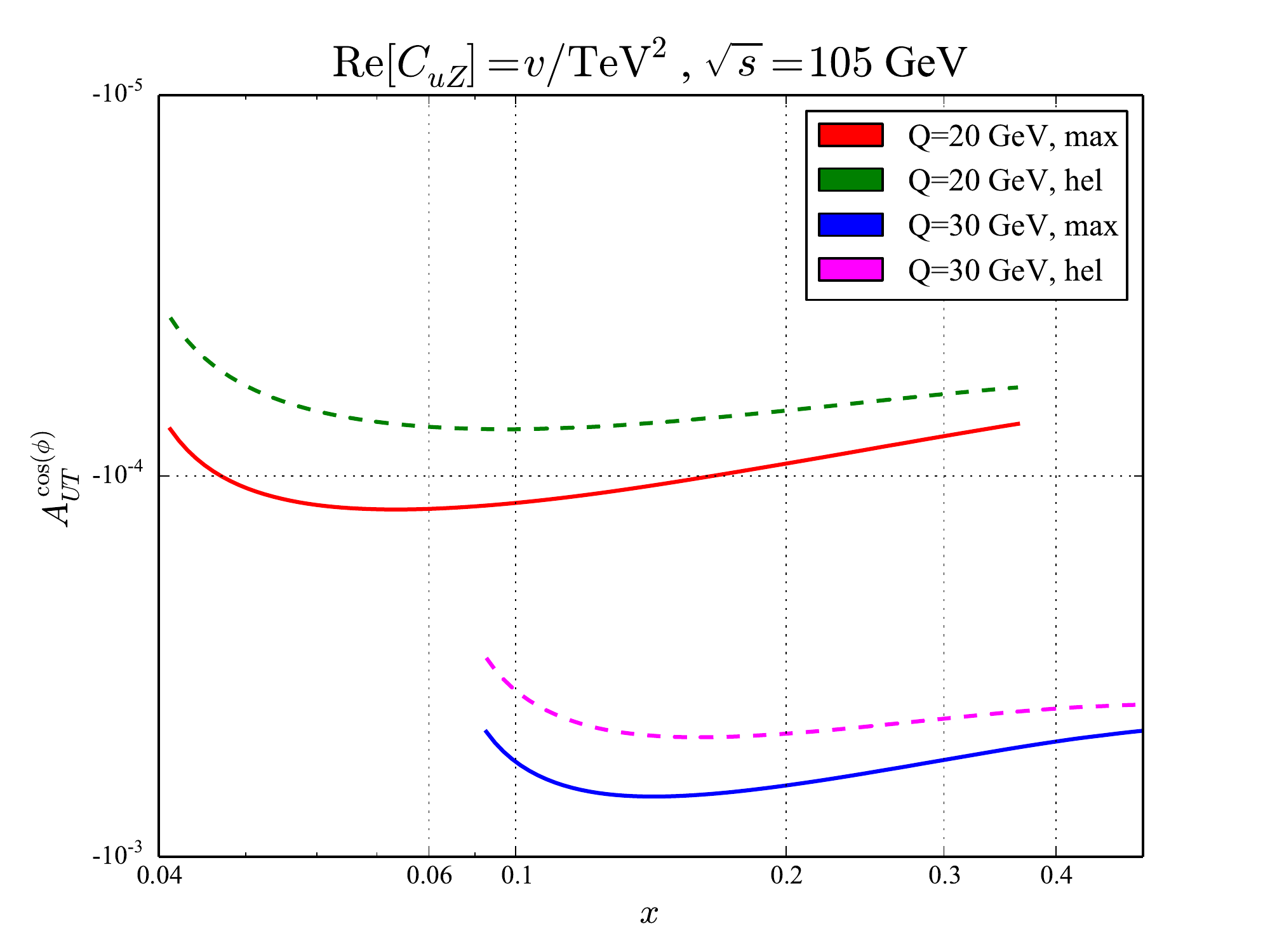}
\caption{The SMEFT contribution to the target asymmetry assuming non-zero $\text{Re}[C_{uZ}]$, for two different scenarios for the transversity distributions.  
\label{fig:ReCuZ}}
\end{figure}

We can perform a similar analysis for the target asymmetry. The target asymmetry probes the up and down quark dipole couplings $C_{u\gamma}$, $C_{uZ}$, $C_{d\gamma}$,  and $C_{dZ}$. This study is complicated by the fact that there are currently only rather poor experimental constraints on the transversity distributions $h_q$. To estimate the effects of potential non-zero values of the quark dipole couplings at the EIC we use model calculations for transversity from~\cite{deFlorian:2017ogw}. Two different model calculations are assumed: a {\it max} scenario in which the transversity distributions saturate the Soffer bound of Eq.~\ref{eq:soffer}, and a {\it helicity} scenario where the transversity distributions are equated to the longitudinal helicity PDFs of Ref.~\cite{deFlorian:2014yva} at a low scale. These two scenarios are meant to represent the two extremes of the possible transversity distribution values. We focus on the real part of $C_{uZ}$ and show results in Fig.~\ref{fig:ReCuZ}. The results for $\text{Re}[C_{u \gamma}]$ are similar in magnitude with the opposite sign. The estimated target asymmetries are slightly smaller than the beam asymmetries. We note that the differences between the two studied transversity distributions are not large. The size of the asymmetries indicate that it may be possible to observe TeV-scale new physics in quark dipole couplings at the EIC, although a quantitative bound on the associated Wilson coefficients will require a determination of the transversity distributions. 

\section{Other experimental constraints}
\label{sec:other}

We review here other experimental constraints on the $C_{f\gamma}$, $C_{fZ}$ couplings in the SMEFT. The dipole couplings to both the quarks and leptons can be probed through the Drell-Yan process at the LHC. The constraints have been studied in~\cite{Boughezal:2021tih}. It is important to note that these contributions to Drell-Yan occur at the sub-leading $1/\Lambda^4$ level in the SMEFT expansion. For non-zero fermion masses there is no interference between the dipole contributions and the SM in Drell-Yan, and therefore the deviation first occurs at the dimension-6 squared level. It is therefore sub-leading compared to dimension-6 vector operators that contribute at $1/\Lambda^2$. This is in contrast to the SSAs studied here, where the dipole terms represent the leading contributions. Assuming that only a single dipole operator contributes at a time, the analysis of~\cite{Boughezal:2021tih} found TeV-scale bounds on linear combinations of the couplings $C_{i\gamma}$ and $C_{iZ}$, where $i=e,q$. We conclude that the potential EIC probes are competitive with those of the Drell-Yan at the LHC and are advantageous from the perspective of new physics interpretation since they represent the leading SMEFT contribution.

There are additionally low-energy constraints on the  dipole couplings, particularly for the electron. The real parts of the Wilson coefficients are probed by measurements of the magnetic moments, while the imaginary parts are strongly constrained by electric dipole moment searches. We note that there is currently an over $5\sigma$ discrepancy between determinations of the electron magnetic moment using either Cesium or Rubidium measurements of the fine structure constant~\cite{Parker_2018,Morel2020}, making this an interesting target for future EIC analyses. We note that the difference is
\be
(\Delta a_e)^{exp-th} = \frac{m_e}{m_{\mu}}\left[ \begin{array}{l} -1.8(7)^{\text{Cs}} \\ 1.0(6)^{\text{Rb}}\end{array} \times 10^{-10}\right].
\label{eq:aeexp}
\ee
A recent analysis of constraints on the $C_{eW}$ and $C_{eB}$ coefficients from magnetic and electric dipole moment measurements can be found in~\cite{Aebischer:2021uvt}. The result for the electron anomalous magnetic moment can be written as
\be
(\Delta a_e)^{SMEFT} = \frac{m_e}{m_{\mu}} \left\{ 2.8\times 10^{-3} C_{eB} - 1.5 \times 10^{-3} C_{eW}\right\} (250 \, \text{GeV})^2.
\ee
Converting these to the $C_{e\gamma}$, $C_{eZ}$ basis using the $\overline{MS}$ values of the weak mixing angle, we find 
\be
(\Delta a_e)^{SMEFT} = \frac{m_e}{m_{\mu}} \left\{ 1.4\times 10^{-3} C_{e\gamma} - 1.3 \times 10^{-5} C_{eZ}\right\} (250 \, \text{GeV}).
\ee
The sensitivity to $C_{eZ}$ is less than $C_{e\gamma}$. This arises because the low-energy theory below the electroweak scale contains only $C_{e\gamma}$. The $C_{eZ}$ dependence is  generated by running above the electroweak scale. Assuming $C_{ei} \sim v/\Lambda^2$, we find that scales of 100 TeV for $C_{e\gamma}$ are needed to match the experiment versus theory difference quoted above in Eq.~(\ref{eq:aeexp}). Scales of  order 10 TeV for $C{eZ}$ are needed to address the difference. We note that the anomalous magnetic moment probes only a single linear combination of $C_{e\gamma}$ and $C_{eZ}$. Using the $y$ dependence of the SSA shown in Eq.~(\ref{ATUsmeft}) both $C_{e\gamma}$ and $C_{eZ}$ can be separately probed at the EIC, making its contribution to the exploration of this sector of the SMEFT important. Although the scale for $C_{e\gamma}$ reachable by the anomalous magnetic moment is beyond what the EIC can probe, the EIC should be able to provide competitive constraints on $C_{eZ}$, especially since the $Z$ dipole contribution can be isolated at the EIC.

\section{Probing the muon anomalous magnetic moment at a muon-ion collider}
\label{sec:muon}

One proposed upgrade for the EIC would replace the initial electron beam with a high-energy muon beam~\cite{Acosta:2021qpx}. In addition to providing a first step toward a TeV-scale muon-muon collider, this machine would extend the physics program of the EIC to include topics such as Higgs physics~\cite{Ahluwalia:2022qsp}. It is possible to achieve a muon polarization reaching 50\% at such a machine with a slight reduction of integrated luminosity~\cite{Acosta:2021qpx}, allowing the muon beam SSA to be measured. Muon beam SSAs are sensitive to the dipole couplings of the muon, $C_{\mu \gamma}$, $C_{\mu Z}$. The real parts of these coefficients are exactly those needed to explain the discrepancy between theory and experiment for the muon anomalous magnetic moment~\cite{Muong-2:2021ojo}, and a muon-ion collider could therefore shed light on this outstanding issue. The imaginary parts of these coefficients lead to a muon electric dipole moment. The current constraints on this quantity are significantly weaker than those on the electron electric dipole moment (EDM)~\cite{Muong-2:2008ebm}. We will find that a muon-ion collider can potentially provide stronger probes of $\text{Im}[C_{\mu Z}]$ than current muon EDM bounds.

\begin{figure}[h!]
\centering
\includegraphics[width=0.7\textwidth]{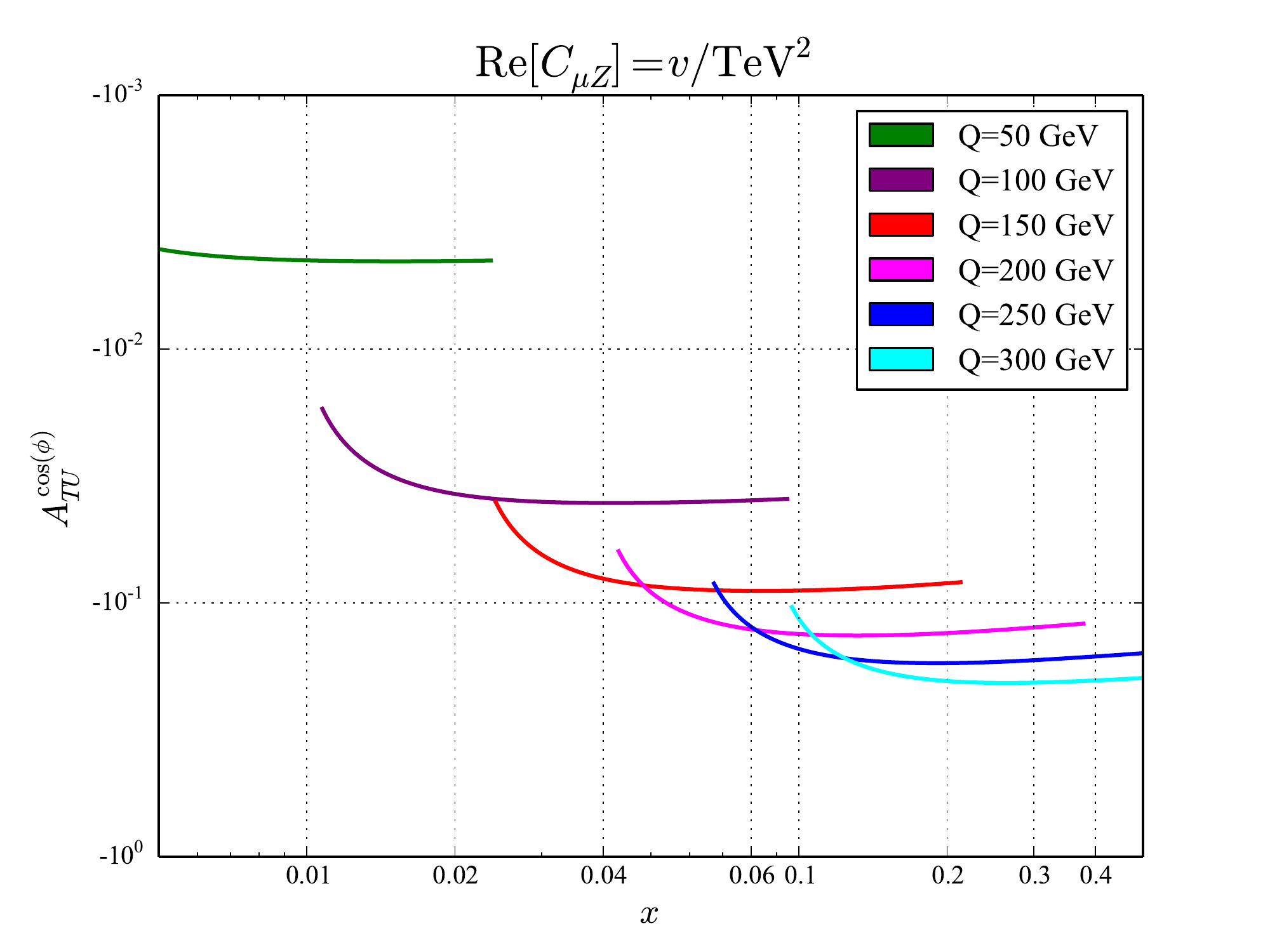}
\caption{The SMEFT contribution to the asymmetry at a muon-ion collider for the real part of  $C_{\mu Z}$ for representative values of $Q$ as a function of Bjorken-$x$. 
\label{fig:ASMEFTmuon}}
\end{figure}

\begin{figure}[h!]
\centering
\includegraphics[width=0.7\textwidth]{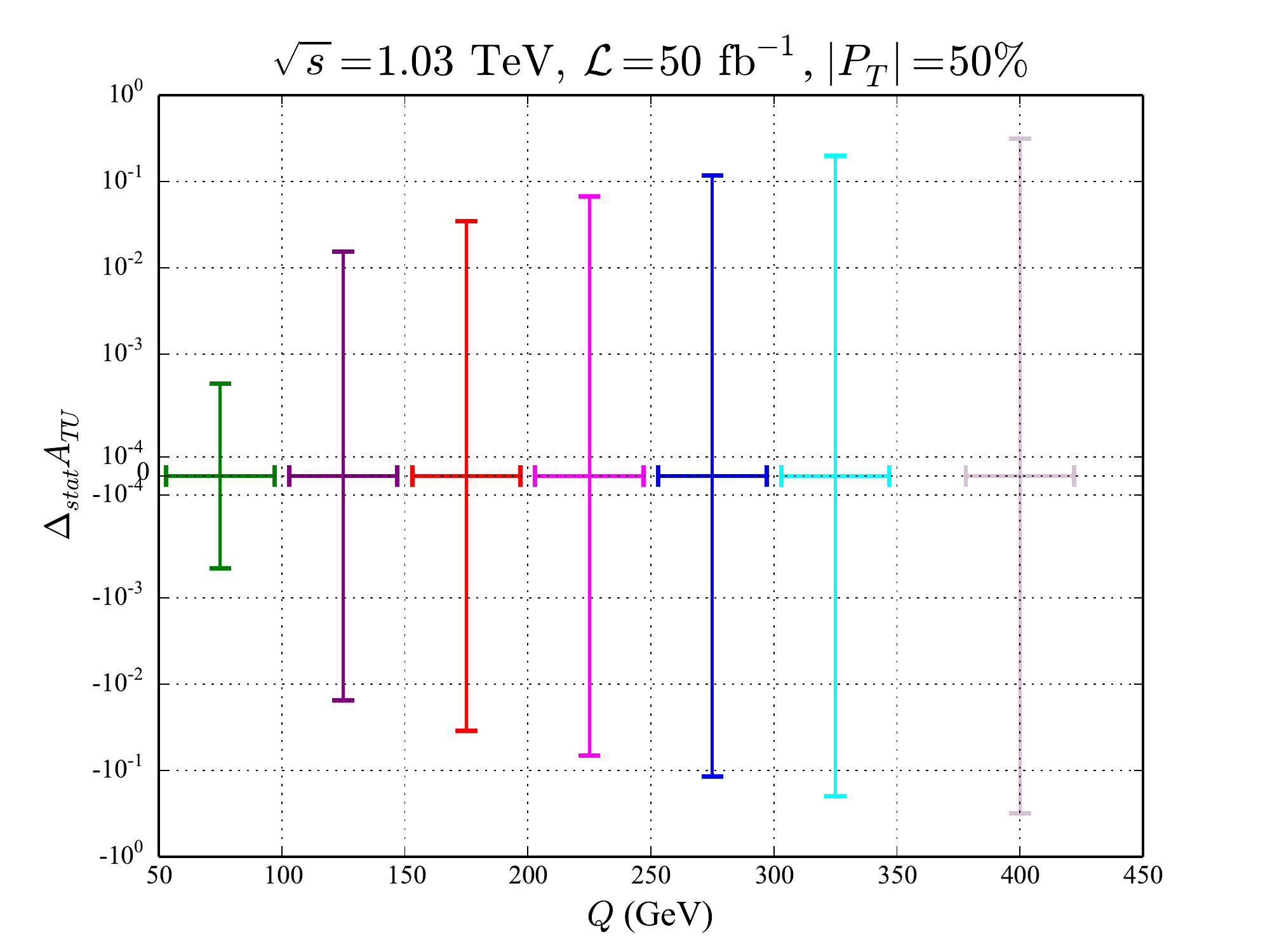}
\caption{The estimated statistical uncertainty on the asymmetry at a future muon-ion collider for $Q$ bins ranging from 50 to 300 GeV. Bins of width 50 GeV are assummed. 
\label{fig:ASMEFTmuonstat}}
\end{figure}

To study the physics potential of beam SSAs to probe muon dipole couplings at a muon-ion collider we assume a 960 GeV muon beam and a 275 GeV proton beam, leading to a center-of-mass energy slightly over 1 TeV. We assume 50\% transverse polarization of the initial muon beam~\cite{Acosta:2021qpx} and 50 fb$^{-1}$ of integrated luminosity. This amount of integrated luminosity is less than the expected 100 fb$^{-1}$, consistent with the expected reduction of luminosity with higher muon polarization. We set $\text{Re}[C_{\mu Z}]=v/\text{TeV}^2$ as before and show the expected SMEFT contribution as a function of $x$ and $Q^2$ in Fig.~\ref{fig:ASMEFTmuon}. The result for $\text{Re}[C_{\mu \gamma}]$ is similar in magnitude with opposite sign. As discussed before, the imaginary parts of the Wilson coefficients give identical contributions to the asymmetry upon replacement of $\text{cos}(\phi) \to \text{sin}(\phi)$ in the weight function. The expected statistical error on the asymmetry given the parameters above is shown in Fig.~\ref{fig:ASMEFTmuonstat}. The asymmetry becomes signifantly larger at a higher energy muon-ion collider, and we expect that scales approaching several TeV can be probed.

To understand the impact of potential muon-ion collider probes of $C_{\mu\gamma}$ and $C_{\mu Z}$ we first recall the analysis of the muon $g-2$ within SMEFT provided in~\cite{Aebischer:2021uvt}. As the momentum transfers at a muon-ion collider will be far above the $Z$-boson mass it is appropriate to compare directly in the SMEFT. Converted into our notation, the result given there for the muon anomalous magnetic moment correction in the SMEFT is
\be
\Delta a_{\mu}^{SMEFT} = 1.1 \times 10^{-3} \left(\frac{\text{Re}[C_{\mu\gamma}]}{1 \, \text{TeV}^{-1}}\right) -1.1 \times 10^{-5} \left(\frac{\text{Re}[C_{\mu Z}]}{1 \, \text{TeV}^{-1}}\right).
\ee
We have assumed a simple leading-order scaling to convert the result of~\cite{Aebischer:2021uvt} at the renormalization scale $\mu=250$ GeV to the $\mu=1$ TeV assumed in the above equation. The effect of higher-order running in this translation is small. The current difference between the theoretical and experimental values is 
\be
\Delta a_{\mu}^{exp-SM} = 251(59) \times 10^{-11}.
\ee
If we assume the scaling $C_{\mu i} = v/\Lambda^2$, and turn on only a single coefficient at a time, we find that scales approaching $\Lambda \approx 300$ TeV for $C_{\mu\gamma}$ are needed to explain the discrepancy, while  $\Lambda \approx 30$ TeV is needed for $C_{\mu Z}$. Both energy scales are beyond the reach of a future muon-ion collider. However, if both coefficients are turned on simultaneously, then the solution to the $\Delta a_{\mu}$ discrepancy can be addressed with $\Lambda \approx 1$ TeV and $C_{\mu\gamma} \approx 0.01 C_{\mu Z}$, which is a suppression of about a loop factor. Although we will not speculate here on the possible origin of such a ratio between the dipole couplings, it is important to probe all possible explanations of the $\Delta a_{\mu}$ discrepancy. A muon-ion collider can test this region of parameter space since it depends upon an entirely different linear combination of the $C_{\mu i}$ than $\Delta a_{\mu}$.

We now study existing constraints on the EDM of the muon, and investigate whether a muon-ion collider can improve upon these bounds. We can again convert the results of~\cite{Aebischer:2021uvt} for the muon EDM into our notation: 
\be
\bigg|\frac{\Delta d_{\mu}}{d_{\mu}^{\,\textrm{exp}}}\bigg| = 7.3 \times 10^2 \left(\frac{\text{Im}[C_{\mu\gamma}]}{1 \, \text{TeV}^{-1}}\right)+1.8 \left(\frac{\text{Im}[C_{\mu Z}]}{1 \, \text{TeV}^{-1}}\right).
\ee
If we again assume the scaling $C_{\mu i} = v/\Lambda^2$ and turn on only a single coefficient at a time, we find  that scales approaching $\Lambda \approx 13$ TeV are probed for  $\text{Im}[C_{\mu\gamma}]$, beyond the reach of the muon-ion collider. However, only $\Lambda \approx 700$ GeV is reached for $\text{Im}[C_{\mu Z}]$. A muon-ion collider can improve upon this constraint.

\begin{figure}[h!]
\centering
\includegraphics[width=0.7\textwidth]{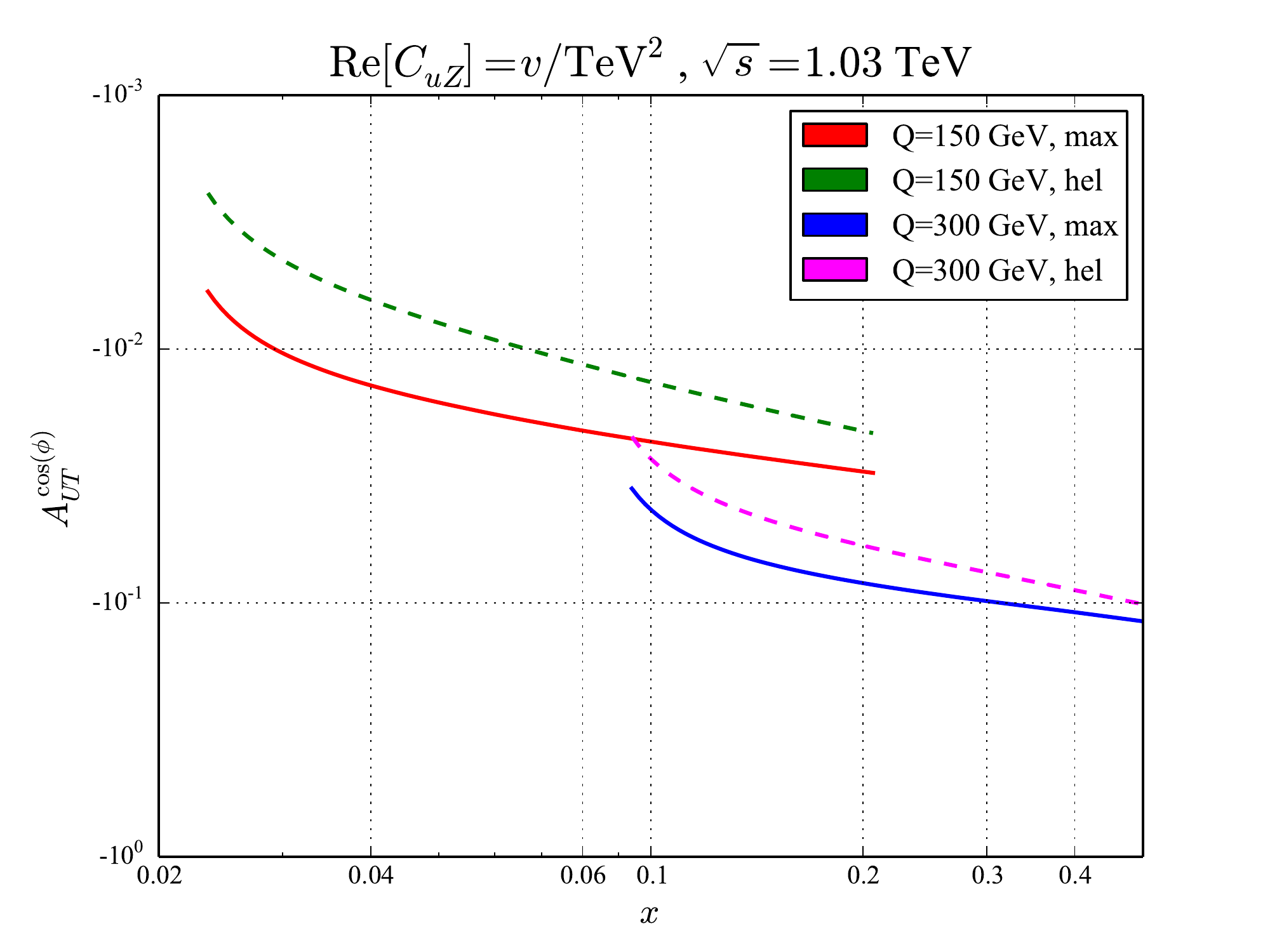}
\caption{The SMEFT contribution to the target asymmetry assuming non-zero $\text{Re}[C_{uZ}]$, for two different scenarios for the transversity distributions, at a future muon-ion collider. 
\label{fig:ReCuZmuon}}
\end{figure}

We can also leverage the higher energy of a proposed muon collider to study the target asymmetry. We show the value of the target asymmetry assuming a non-zero $\text{Re}[C_{uZ}]$ in Fig.~\ref{fig:ReCuZmuon}. The asymmetry is significantly larger than at the nominal EIC, indicating the possibility of significant probes of these Wilson coefficients. To achieve this will require precision extractions of the transversity distributions during the initial running of the EIC, which is possible using semi-inclusive DIS data from polarized EIC collisions~\cite{Gamberg:2021lgx}. 

\section{Conclusions}
\label{sec:conc}

In this manuscript we have studied the potential of transverse SSAs at the EIC to probe electroweak dipole operators of fermions. In the SM these quantities are suppressed by the fermion mass over the momentum transfer, and are much smaller than TeV-scale new physics contributions. We organize potential new physics contributions using the SMEFT. We show that beam SSAs are sensitive to electron dipole couplings to the photon and $Z$-boson, while target asymmetries are sensitive to quark dipole couplings. These couplings are also probed by both high-energy LHC data, and low-energy anomalous magnetic and electric dipole moment measurements. We show that the EIC probes are competitive with the high-energy constraints, and are complementary to the low-energy measurements since they probe different combinations of the new physics couplings. We study the possibility of SSA measurements at a future muon-ion collider. Such an upgrade of the EIC could probe parameter space relevant for the observed discrepancy of the muon anomalous magnetic moment, and could improve on the current muon electric dipole moment limits.

\bigskip
\noindent	{\bf Acknowledgements:} 
	This work originated from discussions at the Center for Frontiers in Nuclear Science (CFNS) workshop 
``Precision QCD predictions for ep Physics at the EIC'', Stony Brook, August 1-5, 2022.
	We thank A.~Manohar and E.~Jenkins for helpful communication regarding~\cite{Aebischer:2021uvt}, and M.~Schlegel for discussions on the 
target spin asymmetry. R.~B. is supported by the US Department of Energy (DOE) contract DE-AC02-06CH11357. The work of D. deF. has been partially supported by ANPCYT and Conicet. F.~P. is supported by the DOE grants DE-FG02-91ER40684 and DE-AC02-06CH11357. 
	 W.~V. is supported by Deutsche Forschungsgemeinschaft (DFG) through the Research Unit FOR 2926 (project 409651613).

\bibliographystyle{h-physrev}
\bibliography{SSA}

\end{document}